\documentclass[prb,twocolumn,superscriptaddress,citeautoscript,showpacs,floats]{revtex4}

\usepackage{amsmath}
\usepackage{amssymb}
\usepackage{epsfig}

\graphicspath{{.}{./EPS/}}

\begin{document}

\title{Features due to spin-orbit coupling in the optical conductivity of single-layer
graphene} 

\author{P. Ingenhoven}
\affiliation{Institute of Fundamental Sciences and MacDiarmid Institute for Advanced
Materials and Nanotechnology, Massey University (Manawatu Campus), Private Bag
11~222, Palmerston North 4442, New Zealand}

\author{J. Z. Bern\'ad}
\affiliation{Institute of Fundamental Sciences and MacDiarmid Institute for Advanced
Materials and Nanotechnology, Massey University (Manawatu Campus), Private Bag
11~222, Palmerston North 4442, New Zealand}

\author{U. Z\"ulicke}
\affiliation{Institute of Fundamental Sciences and MacDiarmid Institute for Advanced
Materials and Nanotechnology, Massey University (Manawatu Campus), Private Bag
11~222, Palmerston North 4442, New Zealand}
\affiliation{Centre for Theoretical Chemistry and Physics, Massey University (Albany
Campus), Private Bag 102904, North Shore MSC, Auckland 0745, New Zealand}

\author{R. Egger}
\affiliation{Institut f\"ur Theoretische Physik, Heinrich-Heine-Universit\"at,
D-40225 D\"usseldorf, Germany}
\date{\today}

\begin{abstract}

We have calculated the optical conductivity of a disorder-free single graphene
sheet in the presence of spin-orbit coupling, using the Kubo formalism. Both
intrinsic and structural-inversion-asymmetry induced types of spin splitting are
considered within a low-energy continuum theory. Analytical results are obtained
that allow us to identify distinct features arising from spin-orbit couplings. We
point out how optical-conductivity measurements could offer a way to determine
the strengths of spin splitting due to various origins in graphene.

\end{abstract}


\maketitle

\section{Introduction}
Graphene is a single sheet of carbon atoms forming a two-dimensional honeycomb
lattice. This material has only recently become available for experimental study, and
its exotic physical properties have spurred a lot of
interest~\cite{novoselov2004,novoselov2005}. Known theoretically since the late
40s~\cite{wallace1947}, graphene is a promising candidate for applications due to
its excellent mechanical properties~\cite{lee2008}, scalability down to nanometer
sizes~\cite{scalability}, and exceptional electronic properties~\cite{castroneto2009}. 
The conical shape of conduction and valence bands near the $K$ and $K^\prime$
points in the Brillouin zone renders graphene an interesting type of quasi-relativistic
condensed-matter system~\cite{zhang2005,novoselov2005_2} where mass-less
Dirac-fermion-like quasiparticles are present at low energy. In contrast to the truly
relativistic case, the spin degree of freedom in their Dirac equation corresponds to
a pseudo-spin that distinguishes degenerate states on two sublattices formed by two
nonequivalent atom sites present in the unit cell. 

The pseudo-spin degeneracy can be broken by spin-orbit interaction (SOI), which 
mixes pseudospin and real spin.  There has been huge interest in SOI in graphene,
resulting in a large body of theoretical~\cite{dressel,ando2000,egger,mele_kane2005,
huertas-hernando2006,min2006,yao2007,boettger2007,zarea,guinea2,guinea3,
rashba2009,fabian2009,stauber2009,kuemmeth2009,gmitra2009} 
and experimental~\cite{kuemmeth2008,laubschat2008,varykhalov2008,rader2009,
varykhalov2009} work. There are two main causes for the SOI in graphene. Firstly,
external electric fields (e.g., due to the presence of a substrate, a backgate, or
adatoms) and local curvature fields (ripples) induce a
SOI~\cite{mele_kane2005,huertas-hernando2006,min2006,rashba2009,gmitra2009}
whose coupling strength we denote by $\Delta_R$.  We refer to this contribution as
the \textit{Rashba\/} SOI in the following. In addition, there is an \textit{intrinsic\/}
SOI~\cite{dressel,mele_kane2005,huertas-hernando2006,min2006,yao2007,
boettger2007,gmitra2009} with strength $\Delta_I$, which is caused by the atomic
Coulomb potentials.

Existence of the intrinsic and Rashba SOIs can be inferred from group-theoretical
arguments~\cite{dressel,mele_kane2005,rashba2009}. However, the actual values
of their respective strengths $\Delta_R$ and $\Delta_I$ are the subject of recent
debate. Initial estimates~\cite{mele_kane2005} have been refined using tight-binding
models~\cite{huertas-hernando2006,min2006} and density-functional
calculations~\cite{yao2007,boettger2007,gmitra2009}. First experimental observations 
of spin-orbit-related effects in graphene's band structure based on ARPES
data~\cite{laubschat2008,varykhalov2008} have later been
questioned~\cite{rader2009,varykhalov2009}. Detailed knowledge about typical
magnitudes and ways to influence $\Delta_R$ and $\Delta_I$ is crucial, e.g., for
understanding spin-dependent transport~\cite{wees2007} and spin-based quantum
devices~\cite{trauzettel2007} in graphene. The desire to identify possible alternative
means of observing, and measuring, spin-orbit coupling strengths in graphene has
provided the motivation for our work reported here. 

We present a theoretical analysis of graphene's optical conductivity
$\sigma(\omega)$, extending previous studies~\cite{ziegler2006,gusynin2006,
ziegler2007,falkovsky2007,peres2008,nicol2008,zhang2008a,gusynin2009,
uz:physe:10}
to the situation with finite SOI. SOI effects on the DC conductivity were investigated in
a recent theoretical study for a bipolar graphene $pn$ junction~\cite{yamakage2009}, 
and the effect of intrinsic SOI on the polarisation-dependent optical absorption of 
graphene was considered in Ref.~~\onlinecite{zhang2008b}. Our study presents the 
analogous scenario for the richer case of the optical conductivity when both intrinsic 
and extrinsic types of SOI are present.  Since $\Delta_R$ can be tuned by external 
fields, we will analyze various situations distinguished by the relative strengths of
$\Delta_R$ and $\Delta_I$.

Our findings suggest that optical-conductivity measurements can be useful to
identify and separate different SOI sources. We work on the simplest theory
level (linear response theory, no interactions, no disorder) and disregard boundary
effects for the moment. The structure of the remainder of this article  is as follows.
In Sec.~\ref{sec:model}, we summarize basics of our calculation of the optical
conductivity based on the Kubo formalism; except for some details that have been
relegated to an Appendix. In Sec.~\ref{sec:results}, we show results for different
relative magnitudes of SOI strengths at finite temperature $T$ and chemical 
potential $\mu$. Finally, in Sec. \ref{sec:summary}, we summarize our results and
discuss their applicability to actual experiments.

\section{Optical conductivity}
\label{sec:model}

We start from a low-energy continuum description of graphene~\cite{castroneto2009},
$H(\mathbf{k})=H_0(\mathbf{k})+H_R+H_I.$ Without the SOI terms $H_R$ and
$H_I$, the single-particle Hamiltonian in plane-wave representation reads
\begin{equation}
H_0(\mathbf{k})= \hbar v \, (k_x \sigma_x +k_y \tau_z \sigma_y), 
\end{equation} 
with Fermi velocity $v\approx 10^6$~m/s. The Pauli matrices $\sigma_{x,y}$ act in
pseudo-spin space, where the two eigenspinors of $\sigma_z$ correspond to
quasiparticle states localized on sites of the $A$ and $B$ sublattice. Analogous
Pauli matrices $\tau_{x,y,z}$ act in the two-valley space spanned by states near the
two $K$ points.  
The part of the effective Hamiltonian describing Rashba SOI is given by
\begin{equation}
H_R=\frac{\Delta_R}{2}\,(\sigma_x \tau_z s_y-\sigma_y s_x),
\end{equation}
where $\Delta_R$ includes both the external electric-field and curvature effects in a
coarse-grained approximation, with the latter assumed to be homogeneous.
The Pauli matrices $s_{x,y,z}$ act in the real spin space. For the intrinsic SOI induced
by atomic potentials, we have
\begin{equation}
H_I=\Delta_I \,\sigma_z  \tau_z s_z.
\end{equation}
The full Hamiltonian is then an $8\times 8$ matrix in the combined sublattice, spin,
and valley space. 

The full Hamiltonian matrix turns out to be block-diagonal in the valley degree of
freedom, and each block can be transformed into the other via a unitary 
transformation. The bulk spectrum -- ignoring subtleties related to the topological
insulator phase encountered for $2\Delta_I>\Delta_R$~\cite{mele_kane2005}
for now -- can then be obtained from a $4\times 4$ Hamiltonian matrix 
in the basis  $(A\uparrow,B\uparrow,A\downarrow,B\downarrow)$ at 
one $K$ point. The valley degree of freedom then merely manifests itself as a
degeneracy factor $g_v=2$.  The energy spectrum is obtained as 
\begin{equation}
\varepsilon_{\mathbf{k},\nu\nu'}=\frac{1}{2} \left( \nu' 
\Delta_R +\nu \sqrt{4 (\hbar v)^2 \mid \mathbf{k} \mid^2 +
(\Delta_R-2\nu'\Delta_I)^2} \right),
\end{equation}
where the combined indices $\nu,\nu'=\pm 1$ label the four bands. The
 corresponding eigenstates 
\begin{equation}
| n\rangle = |\mathbf{k}\rangle \otimes | \nu\nu'\rangle_{\mathbf{k}}
\end{equation}
are composed of a plane wave state 
$\left|\mathbf{k}\right\rangle$ and a $\mathbf{k}$-dependent 4-spinor 
$\left| \nu\nu'\right\rangle_{\mathbf{k}}$.

We compute the  optical conductivity using the standard 
Kubo formula \cite{mandelung},
\begin{equation}
\sigma_{ab}=\int\limits_{-\infty}^0 dt \ e^{i(\omega-i0^+)t} K_{ab}(t),
\end{equation}
where $a,b=x,y$ and the kernel reads
\begin{equation}
K_{ab}=\frac{ie}{\hbar} \mbox{Tr}\left[ e^{-\frac{i}{\hbar} 
H(\mathbf{k}) t} j_a e^{\frac{i}{\hbar} H(\mathbf{k}) t} 
\left[ r_b , \rho_0 \right] \right].
\end{equation}
Here $e$ denotes the electron charge, $r_b$ is a Cartesian component of the
position operator, $\rho_0$ the equilibrium density matrix, and the current operators
are given by
\begin{equation}
\label{eq:currentoperator}
j_a  = \frac{ie}{\hbar} \left[ H(\mathbf{k}), r_a \right]
=\frac{e}{\hbar}\frac{\partial H(\mathbf{k})}{\partial k_a}.
\end{equation} 
Following Ref.~~\onlinecite{ziegler2006}, we use the 
single-particle eigenstates $|n\rangle$ and eigenenergies 
$\varepsilon_n$. The conductivity then reads
\begin{eqnarray}
 \sigma_{ab}(\omega) &\displaystyle = \frac{e^2}{i} \sum\limits_{n,n^\prime} &
\frac{\left\langle n \left| \left[H,r_a \right] \right|n^\prime \right
\rangle\left\langle n^\prime \left| \left[H,r_b \right] \right|n\right\rangle}
{(\varepsilon_{n^\prime}-\varepsilon_n)(\varepsilon_{n^\prime}-
\varepsilon_n+\hbar \omega-i0^+)}\nonumber \\
&& \times \left[ f(\varepsilon_n)-f(\varepsilon_{n^\prime}) \right],
\end{eqnarray}
where $f(\varepsilon)$ is the Fermi-Dirac distribution containing 
the chemical potential $\mu$ and the inverse temperature $\beta=1/(k_B T)$. 

In the absence of a magnetic field, the off-diagonal entries vanish,
$\sigma_{xy}=0$, while symmetry arguments show that $\sigma_{xx}=\sigma_{yy}
\equiv \sigma(\omega)$. At finite $\omega$ in the clean system, only the inter-band
contribution to the conductivity is relevant. Its real part is given by
\begin{eqnarray}
\label{eq:sigma_aa}
&&{\rm Re} \ \sigma(\omega)= 
\pi e^2 \int \frac{d^2 \mathbf{k}}{(2 \pi)^2}{\sum}^\prime
|w^a_{\kappa\nu,\kappa^\prime\nu^\prime}(\mathbf{k})|^2  \nonumber \\
&&  \hspace{1cm} \times
\frac{f(\varepsilon_{\mathbf{k},\kappa\nu})-f(\varepsilon_{\mathbf{k},
\kappa^\prime\nu^\prime})}
{\varepsilon_{\mathbf{k},\kappa^\prime\nu^\prime}-
\varepsilon_{\mathbf{k},\kappa\nu}}
\\ \nonumber
&&\times \left[ \delta(\varepsilon_{\mathbf{k},\kappa\nu}-
\varepsilon_{\mathbf{k},\kappa^\prime\nu^\prime}+\hbar \omega) 
+ \delta(\varepsilon_{\mathbf{k}\kappa^\prime\nu^\prime}
-\varepsilon_{\mathbf{k}\kappa\nu}+\hbar \omega)\right],
\end{eqnarray}
where  
\[
w^a_{\kappa\nu,\kappa^\prime\nu^\prime}(\mathbf{k}) = 
 {}_{\mathbf{k}}\left\langle
\kappa\nu\right| j_a \left|  \kappa^\prime 
\nu^\prime\right\rangle_{\mathbf{k}}
\]
 are the current operator matrix elements in the eigenbasis, 
and ${\sum}^\prime=\sum_{(\kappa\nu)\ne(\kappa^\prime\nu^\prime)}$. 
We also used  
\[
w^a_{\kappa\nu,\kappa^\prime\nu^\prime}(\mathbf{k})=
\left [ w^a_{\kappa^\prime\nu^\prime,\kappa\nu}(\mathbf{k})\right]^\dagger  
\]
since the current operator is Hermitian.
In what follows, we restrict ourselves to the real part of $\sigma(\omega)$
and omit the ``Re'' sign.

The result obtained for $\omega>0$ can be expressed very generally as
\begin{equation} \label{eq:result1}
\frac{\sigma(\omega)}{\sigma_0} =
 2 \pi
\sum\limits_{n=1}^6 F_n(\omega,\Delta_R,\Delta_I,\beta,\mu),
\end{equation}
where $\sigma_0=g_v e^2/(2\pi\hbar)$ and, with the Heaviside function $\Theta$,
the quantities $F_n$ are given by
\begin{widetext}
\begin{eqnarray}
F_1 &=& \tilde{F}_1 \;\; 
\Theta(\hbar \omega - | \Delta_R -2\Delta_I | )\nonumber\\
F_2 &=&\tilde{F}_2\;\;
\Big[ \Theta(\Delta_R-2\Delta_I) \Theta(\hbar \omega - \Delta_R) \Theta(2\Delta_I+\Delta_R-\hbar \omega) +  \Theta(2\Delta_I-\Delta_R) \Theta(\hbar \omega-\Delta_R) \Theta (2 \Delta_R-\hbar \omega)\Big]\nonumber \\
F_3 &=& \tilde{F}_3\;\; 
\Big[ \Theta(\Delta_R-2\Delta_I)+\Theta(2\Delta_I-\Delta_R) \Theta (\hbar \omega-2\Delta_I+\Delta_R)\Big]\nonumber\\
F_4 &=& \tilde{F}_4 \;\;
\Big[\Theta(\Delta_R-2\Delta_I) \Theta(\hbar \omega-2\Delta_R)+\Theta(2\Delta_I-\Delta_R)\Theta(\hbar \omega -2 \Delta_I-\Delta_R) \Big] \nonumber \\
F_5 &=& \tilde{F}_5 \;\;
\Big[ \Theta(\Delta_R-2\Delta_I) \Theta(\hbar \omega-\Delta_R+2\Delta_I) \Theta(\Delta_R-\hbar \omega)+\Theta(2\Delta_I-\Delta_R) \Theta(\Delta_R-\hbar \omega)\Big]\nonumber\\
F_6 &=& \tilde{F}_6 \;\;
\Theta(\hbar \omega-2\Delta_I-\Delta_R).
\label{eq:result2}
\end{eqnarray}
\end{widetext}
The rather lengthy analytical expressions for 
the quantities $\tilde{F}_n (\omega,\Delta_R,\Delta_I,\beta,\mu)$ 
can be found in the Appendix. In Fig.~\ref{f1}, we show the regions in the $\Delta_R
/\hbar \omega - 2\Delta_I/\hbar \omega$-plane where the different $\tilde{F}_n$ 
contribute.
\begin{table}[t]
  \begin{center}
    \begin{tabular}{| c | c | c | c| c | c |}
    \hline
$\tilde{F}_1$ &  $\tilde{F}_2$ &  $\tilde{F}_3$ &  $\tilde{F}_4$ &  $\tilde{F}_5$ &  $\tilde{F}_6$ \\
\hline
    1-3, 5-7, 10, 11 & 6-8 & 1-3, 5-7, 9-11 &  1, 2 & 10-12  & 1, 2, 5 \\
    \hline
    \end{tabular}
  \end{center}
  \caption{\label{t1} List of $\tilde{F}_n$ functions and the regions in which they contribute, as illustrated in Fig.~\ref{f1}.}
\end{table}

\begin{figure}[t]
\begin{center}
\includegraphics[width=2.8in]{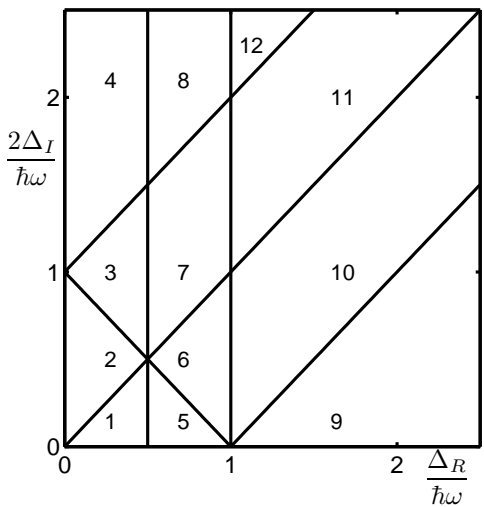}
\caption{\label{f1} Regions in the $\Delta_R/\hbar \omega - 2\Delta_I/\hbar \omega$-plane where the different $\tilde{F}_n$ contribute, cf.\ Table~\ref{t1}. There is no
contribution to $\sigma(\omega)$ from region 4.}
\end{center}
\end{figure}

\section{Results}
\label{sec:results}

We  now discuss the main physical observations arising from Eqs.~(\ref{eq:result1})
and (\ref{eq:result2}).
First, the behavior of the conductivity is qualitatively different in the two regimes
$\Delta_R> 2\Delta_I$ and $\Delta_R< 2\Delta_I$. It is well-known that the latter
regime corresponds to a topological insulator phase while the former yields a
conventional band insulator, with a quantum phase transition in between. For the
topological insulator phase~\cite{mele_kane2005,kane2005_2,brey2006,
bernevig2006}, spin-polarized gapless edge states forming a helical liquid will
dominate the optical conductivity when both $k_B T$ and $\hbar\omega$ are smaller
than the gap energy. In that regime, the conductivity is
expected~\cite{mele_kane2005} to exhibit power-law behavior analogous to that
found for ordinary one-dimensional electron
systems~\cite{giamarchi1988,giamarchi1992}. In what follows, we consider the
frequency and temperature range such that the optical conductivity is still mostly
determined by the bulk states.  

Sharp features are exhibited by the conductivity as a function of frequency $\omega$,
which depend on the relative strength of the two SOI terms and should therefore allow
for a clear identification of these couplings. We start by discussing a few special
cases. For $\Delta_R= 0$ but finite $\Delta_I$, the gapped spectrum consisting of two
doubly (spin-)degenerate dispersion branches leads to a vanishing conductivity for
$\hbar\omega<\Delta_I$, and all other features expected in the presence of a generic
mass gap~\cite{gusynin2006,gusynin2009}. In contrast, for $\Delta_I=0$ but finite
$\Delta_R$, the band structure mimics that of bilayer graphene, only with a gap
smaller by up to 4 orders of magnitude~\cite{dressel,mcclure1957}. The optical
conductivity for this case has the same functional form as the conductivity for bilayer
graphene~\cite{nicol2008,uz:physe:10}, except that the McClure~\cite{mcclure1957} 
interlayer hopping constant is replaced by $\Delta_R$. In particular, it exhibits a
$\delta$-peak at $\hbar\omega=\Delta_R$ and a kink at $\hbar\omega=2\Delta_R$.
With $\hbar v k = \epsilon$, the analytical expression is
\begin{widetext}
 \begin{eqnarray}
&&\frac{\sigma}{\sigma_0}=
\frac{\pi}{2}\delta\left(\hbar \omega-\Delta_R\right)\int_0^{\infty}d\epsilon
\frac{\epsilon\Delta_R}{4\epsilon^2+\Delta_R^2}
 \,\, \left[g\left(\frac{1}{2}\left(\Delta_R+\sqrt{4\epsilon^2 
+\Delta_R^2}\right)\right)
+g\left( \frac{1}{2}\left(\Delta_R-\sqrt{4\epsilon^2 
+\Delta_R^2}\right)\right) \right] \\
&&+ \frac{\pi}{8}g\left(\frac{\hbar \omega}{2}\right)
\left[\frac{\hbar \omega +2\Delta_R}{ 
\hbar \omega+\Delta_R}+\frac{ \hbar \omega -2\Delta_R}{\hbar
 \omega-\Delta_R}\Theta(\hbar \omega-2\Delta_R)\right]
+\frac{\pi}{8}\frac{\Delta_R^2}{(\hbar \omega)^2} 
\left[g\left(\frac{\hbar \omega+\Delta_R}{2}\right)
+g\left(\frac{\hbar\omega-\Delta_R}{2}\right)\right]\Theta(\hbar 
\omega-\Delta_R), \nonumber
\end{eqnarray}
\end{widetext}
where we define the function
\begin{equation}
 g(\varepsilon)=\frac{\sinh (\varepsilon 
\beta)}{\cosh (\mu\beta)+\cosh (\varepsilon \beta)}.
\end{equation}
In the limit $\Delta_R=\Delta_I=0$, the optical conductivity of clean graphene with its
spin-degenerate linear dispersion is recovered~\cite{ziegler2007,falkovsky2007}. The
asymptotic behavior for large frequencies turns out to be independent of the SOI
couplings, with $\sigma$ always approaching the well-known universal value $e^2/ 
(4\hbar)$.

\begin{figure}[t]
\begin{center}
\includegraphics[width=2.8in]{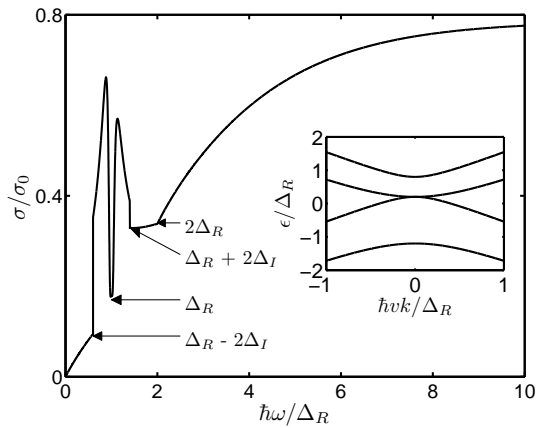}
\caption{\label{f4}
Optical conductivity at $T=1$~K for graphene with $\Delta_R=100\,\mu$eV and
$\Delta_I=0.2\Delta_R$, thus realizing the case $2\Delta_I<\Delta_R$. We set
$\mu=\Delta_I$ to maintain charge neutrality. Inset:  Low-energy part of the
bandstructure. Kinks in the frequency dependence of $\sigma$ arise when new
transitions between different bands become possible at certain critical values of
$\omega$.}
\end{center}
\end{figure}

\begin{figure}[t]
\begin{center}
\includegraphics[width=2.8in]{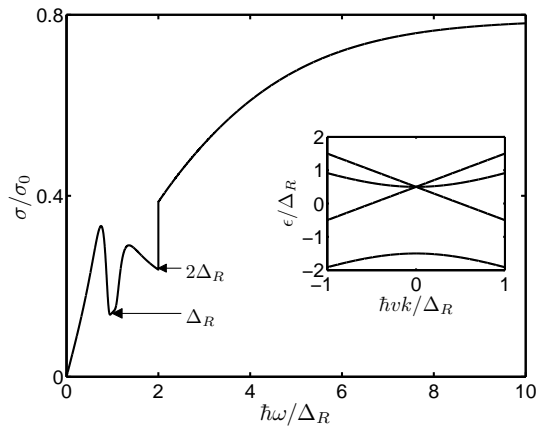}
\caption{\label{f5} 
Optical conductivity at $T=1$~K for graphene obtained for the special case  
$2\Delta_I=\Delta_R$ with $\Delta_R=100\,\mu$eV, setting $\mu=\Delta_I\equiv
\Delta_R/2$ to ensure charge neutrality. Inset: The bandstructure shows that three
bands cross at $k=0$ and, hence, some of the kinks present in Fig.~\ref{f4}
disappear.}
\end{center}
\end{figure}

\begin{figure}[b]
\begin{center}
\includegraphics[width=2.8in]{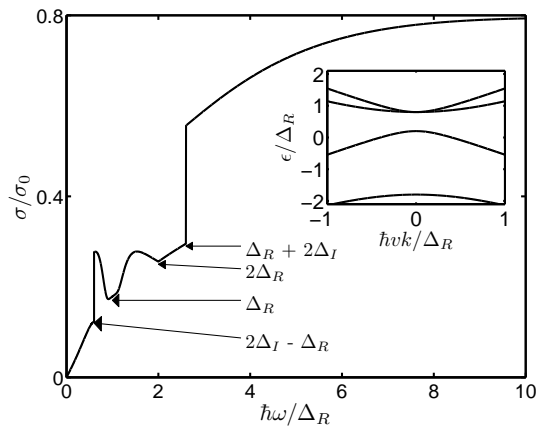}
\caption{\label{f6} 
Same  as Fig.~\ref{f4}, except that $\Delta_I=0.8\Delta_R$ with $\Delta_R=100 \,
\mu$eV, thus realizing the case $2\Delta_I>\Delta_R$. Charge neutrality is maintained 
by setting $\mu=\Delta_R/2$.}
\end{center}
\end{figure}

\begin{figure}[t]
\begin{center}
\includegraphics[width=2.8in]{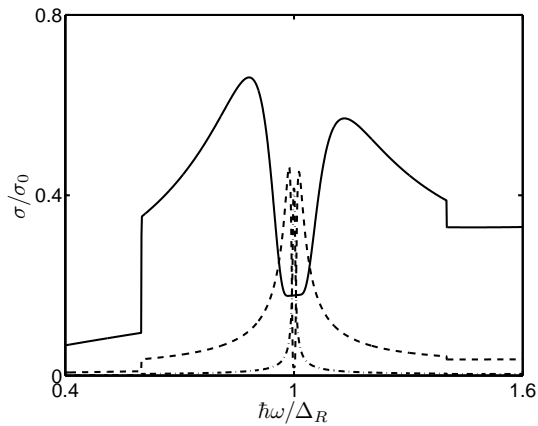}
\caption{\label{f7} 
Same as Fig.~\ref{f4}, focusing on the region $0.4<\hbar\omega/\Delta_R<1.6$.
The solid curve is for $T=1$~K, the dashed curve for $T=10$~K, and the dot-dashed
one for $T=100$~K. The distinct kinks are thermally smeared and suppressed
at elevated temperatures, but  remain visible up to $T\approx 10$~K.}
\end{center}
\end{figure}

The optical conductivity for various situations where both $\Delta_R$ and $\Delta_I$
are finite is shown next in a series of figures. In particular, Fig.~\ref{f4} shows the
case where $\Delta_R>2\Delta_I$. In Fig.~\ref{f5}, we are at the special point
$\Delta_R=2\Delta_I$. Furthermore, Fig.~\ref{f6} illustrates the regime where
$\Delta_R<2\Delta_I$. To be specific, all these figures are for $T=1$~K. Finally,
Fig.~\ref{f7} displays the effects of thermal smearing.

For $2\Delta_I<\Delta_R$, we observe a splitting and widening 
of the $\delta$-peak at $\Delta_R$, while the kink at $2\Delta_R$ 
stays at the same position. In  addition, we observe kinks at 
$\Delta_R\pm2\Delta_I$, see Fig.~\ref{f4}. 
At the quantum phase transition point $2\Delta_I=\Delta_R$, 
the dispersion exhibits a crossing of two massless branches with a massive 
branch, see inset of Fig.~\ref{f5}. As a consequence, certain sharp features
exhibited by the optical conductivity in other cases disappear .
For $2\Delta_I>\Delta_R$, see Fig.~\ref{f6}, the conductivity 
shows kinks at $\hbar\omega=\Delta_R$, at
$\hbar\omega=2\Delta_R$, and at $2\Delta_I\pm\Delta_R$.

We have chosen to show a very wide range of SOI parameters $\Delta_R$ and
$\Delta_I$ in these figures. Previous estimates for these parameters
\cite{mele_kane2005,min2006,huertas-hernando2006,yao2007,boettger2007,
gmitra2009} range from 0.5~$\mu$eV to 100~$\mu$eV for $\Delta_I$, and
0.04~$\mu$eV to 23~$\mu$eV for $\Delta_R$. The Rashba coupling is expected
to be linear in the electric backgate field, with proportionality constant
10~$\mu$eV~nm/V (Ref.~~\onlinecite{gmitra2009}), allowing for an experimental
lever to sweep through a wide parameter range. On the experimental side, the
picture is currently mixed. One recent experimental study~\cite{kuemmeth2008}
finds $\Delta_R=370\,\mu$eV  ($210\,\mu$eV) for electrons (holes) in carbon
nanotubes. A much larger value $\Delta_R=13$~meV has been reported for
graphene sheets fabricated on a nickel surface~\cite{varykhalov2008}. 
 
For low temperatures (e.g., at $T=1$~K in the above figures), the SOI couplings can
be distinguished by the different peak structures appearing in the optical conductivity. 
Increasing the temperature leads to thermal smearing of those features, as illustrated
in Fig.~\ref{f7}. However, the characteristic  SOI-induced peak and kink features
should still be visible in the optical conductivity up to $T\approx 10$~K, albeit with a
smaller magnitude.

\section{Conclusions}
\label{sec:summary}

We have calculated the optical conductivity for a graphene monolayer
including the two most relevant spin-orbit couplings, namely
the intrinsic atomic contribution $\Delta_I$ and 
the curvature- and electric-field-induced Rashba term $\Delta_R$. 
Our result for the optical conductivity, which we presented 
for finite temperature and chemical potential, shows kinks and/or
peaks at frequencies corresponding to $\Delta_R$, $2\Delta_R$,
 and $|\Delta_R\pm 2\Delta_I|$. 
Measuring the optical conductivity in a frequency range
covering these energy scales can be expected to yield detailed insights
into the nature of spin-orbit interactions in graphene.

We did not analyze disorder effects but expect all sharp features to broaden
since the $\delta$-functions in Eq.~(\ref{eq:sigma_aa}) effectively become
Lorentzian peaks. We also did not consider the effect of electron-electron
interactions. While renormalization group studies indicate that weak unscreened
interactions are marginally irrelevant~\cite{castroneto2009}, interactions may still
play an important role. For instance, Ref.~~\onlinecite{grushin2009} considers
interaction effects on the optical properties of doped graphene without spin-orbit
coupling. Interactions cause inter-band (optical) and intra-band (Drude) transitions
and thus a  finite DC conductivity. We expect that the peak and kink structures
arising from the spin-orbit  couplings survive, however, because the relevant
contributions are additive.

Recent experimental studies suggest that an optical measurement 
of the conductivity in the energy range relevant for SOI should be possible.
Fei \textit{et al.}~\cite{fei2008} have measured the optical conductivity from
$\hbar \omega=1.54$~eV up to $4.13$~eV. Slightly lower energies ($0.2$~eV to 
$1.2$~eV) were reached in Ref.~~\onlinecite{mak2008}. We suggest to perform
low-temperature experiments at microwave frequencies, with energies ranging from
several $\mu$eV to a few meV. 

\acknowledgments

Useful discussions with M.~J\"a\"askel\"ainen are gratefully acknowledged. JZB is 
supported by a postdoctoral fellowship from the Massey University Research Fund.
Additional funding was provided by the German Science Foundation (DFG) through
SFB Transregio 12.

\appendix
\section{Definition of auxiliary functions}
\label{appendixA}

Here we provide the six functions $\tilde{F}_n (\omega,\Delta_R,\Delta_I,\beta,\mu)$ (with $n=1,\ldots,6$) entering Eq.~(\ref{eq:result2}). 
We use the following abbreviations:
\begin{eqnarray*}
\epsilon_1(y)&=&\frac{1}{2} \left(\Delta_R -\sqrt{(\Delta_R -2 \Delta_I )^2+4 y ^2}\right),\\
\epsilon_2(y)&=&\frac{1}{2} \left(\Delta_R +\sqrt{(\Delta_R -2 \Delta_I)^2+4 y^2}\right),\\
\epsilon_3(y)&=&\frac{1}{2} \left(-\Delta_R -\sqrt{(\Delta_R+2 \Delta_I )^2+4 y^2}\right),\\
\epsilon_4(y)&=&\frac{1}{2} \left(-\Delta_R +\sqrt{(\Delta_R +2 \Delta_I)^2+4 
y^2}\right).
\end{eqnarray*}
Furthermore, we define the quantities (setting here $\hbar=1$ for simplicity)
\begin{widetext}
\begin{eqnarray*}
y_1&=&\frac{1}{2} \sqrt{-4 \Delta_I ^2+4 \Delta_I 
 \Delta_R -\Delta_R ^2+\omega ^2},
\\
y_2&=&\frac{\sqrt{\omega } \sqrt{8 \Delta_I^2 \Delta_R -2 
\Delta_R^3-4 \Delta_I^2 \omega +5 \Delta_R^2 \omega -4 \Delta_R  
\omega ^2+\omega ^3}}{\sqrt{4 \Delta_R ^2-8 \Delta_R  \omega +4 \omega ^2}} ,
\\
y_3&=&\frac{\sqrt{-8 \Delta_I^2 \Delta_R  \omega +2 \Delta_R^3 \omega -4 \Delta_I^2 \omega ^2+5 \Delta_R^2 \omega ^2+4 \Delta_R  
\omega ^3+\omega ^4}}{2 \sqrt{\Delta_R ^2+2 \Delta_R  \omega +\omega ^2}},
\\
y_4&=&\frac{\sqrt{\omega } \sqrt{8 \Delta_I^2 \Delta_R -2 \Delta_R ^3
-4 \Delta_I^2 \omega +5 \Delta_R^2 \omega -4 \Delta_R
  \omega ^2+\omega ^3}}{\sqrt{4 \Delta_R ^2-8 \Delta_R  \omega +4 \omega ^2}},
\\
y_5&=&\frac{\sqrt{\omega } \sqrt{8 \Delta_I^2 \Delta_R -2
 \Delta_R^3-4 \Delta_I^2 \omega +5 \Delta_R^2 \omega -4 \Delta_R 
 \omega ^2+\omega ^3}}{\sqrt{4 \Delta_R ^2-8 \Delta_R  \omega +4 \omega ^2} },
\\
y_6&=&\frac{1}{2} \sqrt{-4 \Delta_I^2-4 \Delta_I  \Delta_R -\Delta_R^2+\omega ^2}.
\end{eqnarray*}
Finally, we define $\Delta_\pm=\Delta_R \pm 2\Delta_I$.
With these conventions,
the functions  $\tilde{F}_n (\omega,\Delta_R,\Delta_I,\beta,\mu)$ 
can be expressed as follows:
\begin{eqnarray*}
\tilde{F}_1 &=& [f(\epsilon_1(y_1))-f(\epsilon_2(y_1))]
\frac{y_1 \Delta_-^2}{16 \left(4 y_1^2+\Delta_-^2\right)^{3/2} }\left
|\frac{\sqrt{4 y_1^2+\Delta_-^2}}{y_1}\right|,
\\ \tilde{F}_2 &=& [f(\epsilon_1(y_2))-f(\epsilon_3(y_2))]
{\left|\frac{4 y_2}{\sqrt{4 y_2^2+\Delta_-^2}}-\frac{4 y_2}{\sqrt{4 y_2^2+\Delta_+^2}}\right|^{-1}}
\times\\ 
&&\frac{y_2^3 \left(-2 \Delta_R +\sqrt{4 y_2^2+\Delta_-^2}-\sqrt{4 y_2^2+\Delta_+^2}\right)}
{\left(4 y_2^2+\Delta_-  
\left( \Delta_- -\sqrt{4 y_2^2+\Delta_-^2 }\right)\right) 
\left(4 y_2^2+\Delta_+ \left(\Delta_+ +\sqrt{4 y_2^2+\Delta_+^2}\right)\right)},
\\
\tilde{F}_3 &=& [f(\epsilon_1(y_3))-f(\epsilon_4(y_3))]\times\\
&&\frac{y_3^2 \sqrt{4 y_3^2+\Delta_-^2} \sqrt{4 y_3^2+\Delta_+^2} \left(-2 \Delta_R +\sqrt{4 y_3^2+\Delta_-^2}+\sqrt{4 y_3^2+\Delta_+^2}\right)}{4 \left(\sqrt{4 y_3^2+\Delta_-^2}+\sqrt{4 y_3^2+\Delta_+^2}\right) 
\left(4 y_3^2+\Delta_- \left(\Delta_- -\sqrt{4 y_3^2+\Delta_-^2}\right)\right) 
\left(4 y_3^2+\Delta_+  \left(\Delta_+ -\sqrt{4 y^2+\Delta_+^2}\right)\right)},
\\
\tilde{F}_4 &=& [f(\epsilon_2(y_4))-f(\epsilon_3(y_4))]\times\\
&&\frac{y_4^2 \sqrt{4 y_4^2+\Delta_-^2} \sqrt{4 y_4^2+\Delta_+^2} \left(2 \Delta_R +\sqrt{4 y_4^2+\Delta_-^2}+\sqrt{4 y_4^2+\Delta_+^2}\right)}
{4 \left(\sqrt{4 y_4^2+\Delta_-^2}+\sqrt{4 y_4^2+\Delta_+^2}\right) \left(4 y_4^2+\Delta_-  \left(\Delta_- +\sqrt{4 y_4^2+\Delta_-^2}\right)\right) 
\left(4 y_4^2+\Delta_+ \left(\Delta_+ +\sqrt{4 y_4^2+\Delta_+^2}\right)
\right)},
\\
\tilde{F}_5 &=& [f(\epsilon_2(y_5))-f(\epsilon_4(y_5))]
\left|\frac{4 y_5}{\sqrt{4 y_5^2+\Delta_-^2}}-\frac{4 y_5}{\sqrt{4 y_5^2+\Delta_+^2}}\right|^{-1}\times\\
&&\frac{y_5^3 \left(2 \Delta_R +\sqrt{4 y_5^2+\Delta_-^2}-\sqrt{4 y_5^2+\Delta_+^2}\right)}
{\left(4 y_5^2+\Delta_- \left(\Delta_- +\sqrt{4 y_5^2+\Delta_-^2}\right)\right) 
\left(4 y_5^2+\Delta_+ \left(\Delta_+ -\sqrt{4 y_5^2+\Delta_+^2}\right)\right)},
\\
\tilde{F}_6 &=& [f(\epsilon_3(y_6))-f(\epsilon_4(y_6))]
\frac{y_6 \Delta_+^2}{16 \left(4 y_6^2+\Delta_+^2\right)^{3/2}}\left|\frac{\sqrt{4 y_6^2+\Delta_+^2}}{y_6}\right|.
\end{eqnarray*}
\end{widetext}

\end{document}